\documentclass[preprint,aps]{revtex4}
\newcommand{\be}{\begin{equation}}
\newcommand{\ee}{\end{equation}}
\newcommand{\ba}{\begin{eqnarray}}
\newcommand{\ea}{\end{eqnarray}}

\begin{document}

\draft

\title{Brans-Dicke Scalar Field as a Unique k-essence} 
       
\author{Hongsu Kim\footnote{e-mail : hongsu@astro.snu.ac.kr}}

\address{Astronomy Program, SEES, Seoul National University, Seoul, 151-742, KOREA}

%\date{November, 2004}

\begin{abstract}
In the present work, the Brans-Dicke theory of gravity is taken as a possible theory of
k-essence coupled to gravity and then the role played by the Brans-Dicke scalar field in relation to
the unified model for dark matter and dark energy is explored. It has been realized that the BD scalar 
field does indeed play a role of a k-essence, but in a very unique way which distinguishes it 
from other k-essence fields studied in the literature such as the (generalized) Chaplygin gas or the 
purely kinetic k-essence studied recently by Scherrer. That is, instead of providing an interpolation 
between the dark matter-like evolution and the dark energy-like evolution, the BD theory
predicts the emergence of a yet-unknown {\it zero acceleration} epoch which is an intermediate stage 
acting as a ``crossing bridge'' between the decelerating matter-dominated era and the accelerating phase.

\end{abstract}


\maketitle

\narrowtext
%\twocolumn

\newpage
\begin{center}
{\rm\bf I. Introduction}
\end{center}

Perhaps one of the greatest challenges in the theoretical cosmology today would be to understand
the emergence and nature of the observed late-time acceleration of the universe and provide an answer
to the so-called ``cosmic coincidence conundrum'' which concerns the puzzle :
why at the present epoch are the energy densities of dark energy and of dust-like dark matter
of the same order of magnitude ? For instance, according to the recent WMAP data \cite{wmap}, the universe
energy density appears to consist of approximately 4-per cent of that of visible matter, 21-per cent of that of
dark matter and 75-per cent of that of dark energy. Up until now, the most conservative candidate for the
dark energy is the cosmological constant $\Lambda$ and perhaps the most fashionable candidates with
non-trivial dynamics are quintessence and k-essence. Particularly, the k-essence models \cite{kessence} 
employ rather exotic scalar fields with non-canonical (non-linear) kinetic terms which typically lead to 
the {\it negative} pressure. Usually the emergence of such exotic 
type of scalar fields with non-canonical kinetic terms has been attributed to the string/supergravity
theories in which the non-linear kinetic terms generically appear in the effective action describing
moduli and massless degrees of freedom. In this regard, here we focus particularly on the Brans-Dicke
(BD) theory of gravity \cite{bd} (which is one of the simplest extensions/modifications of Einstein's general
relativity) as it involves probably the simplest form of such non-linear kinetic term for the BD scalar
field. Besides, the BD scalar field (and the BD theory itself) is not of quantum origin. Rather it is
classical in nature and hence can be expected to serve as a very relevant candidate to play some role in 
the late-time evolution of the universe such as the present epoch.
Therefore in our present study, the BD theory is employed as a theory of
{\it k-essence field} (with non-canonical kinetic term) coupled {\it non-minimally} to gravity.
Then in this context, unlike the ``scalar-tensor theory'' spirit of the original BD gravity, 
the BD scalar field is not viewed as a (scalar) part of the gravitational degrees of freedom but
instead is thought of as playing the role of a k-essence which is a matter degree of freedom.

\begin{center}
{\rm\bf II. BD scalar field as a k-essence}
\end{center}

The spirit of Brans-Dicke extension of general relativity is an attempt to properly incorporate the
Mach's principle \cite{weinberg} and Dirac's large number hypothesis \cite{weinberg} in which Newton's
constant is allowed to vary with space and time.
In general, the Brans-Dicke theory of gravity, in the presencs of matter with Lagrangian 
$\mathcal{L}_{M}$, is described by the action 
\begin{eqnarray}
S = \int d^4x \sqrt{g}\left[{1\over 16\pi}\left(\Phi R - \omega {{\nabla_{\alpha}\Phi
\nabla^{\alpha}\Phi }\over \Phi}\right) + \mathcal{L}_{M}\right]
\end{eqnarray}
where $\Phi $ is the BD scalar field representing the inverse of Newton's constant which is allowed to
vary with space and time and $\omega $ is the generic dimensionless parameter of the
theory. Extremizing this action then with respect to the metric $g_{\mu \nu}$, the
BD scalar field $\Phi $ yields the classical field equations given respectively by 
\begin{eqnarray}
G_{\mu \nu} &=& R_{\mu \nu} - {1\over 2}g_{\mu \nu}R = 8\pi T^{BD}_{\mu \nu}
+ {8\pi \over \Phi}T^{M}_{\mu \nu}, ~~~{\rm where} \\
T^{BD}_{\mu \nu} &=& {1\over 8\pi}\left[{\omega \over \Phi^2}(\nabla_{\mu}\Phi \nabla_{\nu}\Phi
- {1\over 2}g_{\mu \nu}\nabla_{\alpha}\Phi \nabla^{\alpha}\Phi) + {1\over \Phi}(\nabla_{\mu}
\nabla_{\nu}\Phi - g_{\mu \nu}\nabla_{\alpha}\nabla^{\alpha}\Phi)\right], \nonumber \\
T^{M}_{\mu \nu} &=& P_{M}g_{\mu\nu} + (\rho_{M}+P_{M})U_{\mu}U_{\nu} ~~~{\rm and} \nonumber  \\
\nabla_{\alpha}\nabla^{\alpha}\Phi &=& {8\pi \over (2\omega + 3)}T^{M\lambda}_{\lambda} 
= {8\pi \over (2\omega + 3)}(3P_{M}-\rho_{M}).
\end{eqnarray}
Note that $T^{M}_{\mu\nu}=(2/\sqrt{g})\delta (\sqrt{g}\mathcal{L}_{M})/\delta g^{\mu\nu}$, but
$T^{BD}_{\mu\nu}$ cannot be defined in a similar manner. This is due to the presence of the 
non-minimal coupling term $\sim \sqrt{g}\Phi R$ in the action which obscures the separation of
the scalar ($\Phi $) Lagrangian from the tensor ($g_{\mu\nu}$) Lagrangian. Indeed, the expression for
$T^{BD}_{\mu\nu}$ and the coefficient factor $(8\pi /\Phi )$ in front of $T^{M}_{\mu\nu}$ originate
from this non-minimal coupling term $\sim \sqrt{g}\Phi R$ upon extremizing the total action $S$ in
eq.(1) with respect to $g_{\mu\nu}$. Therefore, we do not have any freedom/arbitrariness in defining
$T^{BD}_{\mu\nu}$ in an alternative form.
Next, $P_{M}$ and $\rho_{M}$ are the pressure and the energy density of matter consisting of those of 
radiation (denoted by $rad $) and dust (denoted by $m$), i.e., 
$P_{M}=P_{rad}+P_{m}=P_{rad}$ and $\rho_{M}=\rho_{rad}+\rho_{m}$ respectively. In this perfect
fluid approximation for matter, $U^{\alpha}=dX^{\alpha}/d\tau $ (where $\tau $ denotes the proper
time) is defined to be the 4-velocity of a perfect fluid element normalized such that
$U^{\alpha}U_{\alpha}=-1$. And here we would like to stress that we shall work in the context of 
original BD theory format not some conformal transformation of it. That is, we shall work in the 
Jordan frame, not in the Einstein frame.
Note that it is the original spirit \cite{bd} of BD theory of gravity in which the BD scalar
field $\Phi$ is prescribed to remain strictly massless by forbidding its direct interaction
with matter fields. Note also that the pure BD theory can be thought of as a kind of 
theory of k-essence (played by the BD scalar field $\Phi $) 
having a {\it non-canonical} (or non-linear, non-standard, ...) kinetic term and being coupled to 
gravity non-minimally. Then one consequence of this is the fact that the statement 
such as ``$\omega \to \infty$ amounts to the Einstein gravity limit'' loses meaning and 
the BD parameter $\omega $ remains a (yet) completely undetermined parameter of the theory. \\
We now work in the spatially-flat ($k=0$) Friedman-Robertson-Walker (FRW) metric by assuming the
homogeneity and isotropy of the spacetime, i.e.,
\begin{eqnarray}
ds^2 = -dt^2 + a^2(t)[\frac{dr^2}{1-kr^2} + r^2(d\theta^2 + \sin^2 \theta d\phi^2)], 
~~~\Phi = \Phi(t)
\end{eqnarray}
where $k=0$ for the spatially-flat case and $a(t)$ denotes the scale factor. Then in terms of this
FRW metric, the Friedmann equation representing the Einstein equation and the Euler-Lagrange's equation
of motion for the BD scalar field read, respectively
\begin{eqnarray}
\left(\frac{\dot{a}}{a}\right)^2 = \frac{\omega}{6}\left(\frac{\dot{\Phi}}{\Phi}\right)^2
- \left(\frac{\dot{a}}{a}\right)\left(\frac{\dot{\Phi}}{\Phi}\right), 
~~~\ddot{\Phi} + 3\left(\frac{\dot{a}}{a}\right)\dot{\Phi} = 0.
\end{eqnarray}
We now attempt to solve these coupled, non-linear field equations. To this end, we start by assuming
the solution ansatz given by
\begin{eqnarray}
\Phi(t) = \frac{1}{G_{0}}a^{n}(t), ~~~a(t) = a(0)\left[1 + \frac{\beta}{\sqrt{G_{0}}}t\right]^{\alpha}
\end{eqnarray}
where $G_{0}$ denotes the present value of Newton's constant from which the {\it effective} Newton's 
constant, i.e., the inverse of the BD scalar field deviates and $\beta$ is some dimensionless parameter.
Note that (neglecting all matter contents) $G_{0}$ is the only fundamental scale in this theory.
First, substituting the first of this solution ansatz into the Friedmann equation in (5) yields the 
algebraic equation for the power index $n$ as $\omega n^2 - 6n - 6 =0$ which in turn gives
$n = (3 \pm \sqrt{6\omega + 9})/\omega $. Next, substituting the same ansatz this time into the BD scalar
field equation in (5) leads to a non-linear equation for only the scale factor,
$\left(\ddot{a}/a\right) + (n+2)\left(\dot{a}/a\right)^2 = 0.$
Finally, substituting the second of the solution ansatz into this equation allows us to determine the
power $\alpha $ as $\alpha = 1/(n+3)$. 
Note that in addition, the field equations in eq.(5) should be supplemented by the 
energy-momentum conservation, $\nabla_{\nu}T^{\mu\nu}_{BD} = 0$. Treating the BD scalar field as a
perfect fluid, i.e.,
$\tilde{T}^{BD}_{\mu\nu} = T^{BD}_{\mu\nu}/G_{0} = P_{BD}g_{\mu\nu} + (\rho_{BD}+P_{BD})U_{\mu}U_{\nu}$,
$\nabla_{\nu}T^{\mu\nu}_{BD} = 0$ is given, in terms of the FRW-metric, by
\begin{eqnarray}
\dot{\rho}_{BD} = -3\left(\frac{\dot{a}}{a}\right)(\rho_{BD}+P_{BD}).
\end{eqnarray}
Note also that this energy-momentum conservation is indeed a consistency condition originating from
the (geometric) Bianchi identity $\nabla_{\nu}G^{\mu\nu}=0$ with $G_{\mu\nu}$ being the Einstein 
tensor given above in eq.(2). 
Therefore, we now need to check if the general solution in eq.(6) satisfies the energy conservation
equation given in eq.(7).
Thus to this end, we begin with the evaluation of the energy density and the pressure
of the BD scalar field. Thus from  
$\tilde{T}^{BD ~\mu}_{\nu} = P_{BD}\delta^{\mu}_{\nu} + (\rho_{BD}+P_{BD})U^{\mu}U_{\nu}$,
we can read off the energy density and the pressure as
\begin{eqnarray}
\rho_{BD} = - \tilde{T}^{t}_{t} = \tilde{T}_{tt},  
~~~P_{BD} = \tilde{T}^{r}_{r} = \tilde{T}^{\theta}_{\theta} = \tilde{T}^{\phi}_{\phi}.
\end{eqnarray}
Note also that the pressure can be calculated using an alternative expression,
$P_{BD} = (\tilde{T}^{\lambda}_{\lambda} + \tilde{T}_{tt})/3$. Thus using eq.(8) and the expression for the energy-momentum 
tensor for the BD scalar field given in eq.(2), we get
\begin{eqnarray}
\rho_{BD} &=& \frac{1}{16\pi G_{0}}\left[\omega \left(\frac{\dot{\Phi}}{\Phi}\right)^2 - 6 \left(\frac{\dot{a}}{a}\right)
\left(\frac{\dot{\Phi}}{\Phi}\right)\right], \\
P_{BD} &=& \frac{1}{16\pi G_{0}}\left[\omega \left(\frac{\dot{\Phi}}{\Phi}\right)^2 + 4 \left(\frac{\dot{a}}{a}\right)
\left(\frac{\dot{\Phi}}{\Phi}\right) + 2\left(\frac{\ddot{\Phi}}{\Phi}\right)\right]. \nonumber
\end{eqnarray}
Next, from the expressions for exact solutions eq.(6), it follows that
\begin{eqnarray}
\left(\frac{\dot{\Phi}}{\Phi}\right) &=& n\left(\frac{\dot{a}}{a}\right) =
\left(\frac{n}{n+3}\right)\tilde{\beta}\left[\frac{a(t)}{a(0)}\right]^{-(n+3)}, \\
\left(\frac{\ddot{\Phi}}{\Phi}\right) &=&  n\left[\left(\frac{\ddot{a}}{a}\right) + (n-1)
\left(\frac{\dot{a}}{a}\right)^2\right] = -\frac{3n}{(n+3)^2}
\tilde{\beta}^2\left[\frac{a(t)}{a(0)}\right]^{-2(n+3)} \nonumber
\end{eqnarray}
where $\tilde{\beta} = \beta /\sqrt{G_{0}}$.
Therefore from eqs.(7), (9) and (10), the energy-momentum conservation equation now gives
\begin{eqnarray}
\omega = -\frac{3}{2}, ~~~n = - 2
\end{eqnarray}
and hence finally the solution reads
\begin{eqnarray}
a(t) = a(0)\left[1 + \frac{\beta}{\sqrt{G_{0}}}t\right], 
~~~\Phi(t) = \frac{1}{G_{0}}a^{-2}(0)
\left[1 + \frac{\beta}{\sqrt{G_{0}}}t\right]^{-2}. 
\end{eqnarray}
Note also that this set of values in eq.(11) is consistent with 
$n = (3 \pm \sqrt{6\omega + 9})/\omega $ that has been determined from the field equations. 
Next, it is noteworthy that the field equations leave the BD 
$\omega $-parameter and the power index $n$ undetermined but it is the energy-momentum conservation (which
is a consistency condition) that finally determines their values. Having obtained the explicit expressions
for energy density and pressure of the BD scalar field (being treated as a perfect fluid), we now
attempt to determine the equation of state and the speed of sound. Firstly, using eqs.(9), (10) and (11),
we have
\begin{eqnarray}
\rho_{BD} = \frac{1}{16\pi G_{0}}\left[6\tilde{\beta}^2 \left(\frac{a(t)}{a(0)}\right)^{-2}\right],
~~~P_{BD} = - \frac{1}{16\pi G_{0}}\left[2\tilde{\beta}^2 \left(\frac{a(t)}{a(0)}\right)^{-2}\right]
\end{eqnarray}
Then from eq.(13), one can next determine the equation of 
state to be $w_{BD} = P_{BD}/\rho_{BD} = -2\tilde{\beta}^2/6\tilde{\beta}^2 = -1/3$ 
(and hence, for later use, note that $(\rho_{BD} + 3P_{BD}) = 0$), namely, $P_{BD} = -\rho_{BD}/3$
meaning that the BD scalar field turns out to be a {\it barotropic} (i.e., constant $w$) perfect fluid.
Next, the computation of the sound speed needs rather careful treatment. Namely using again eqs.(9) and (10),
it follows that
\begin{eqnarray}
\dot{\rho}_{BD}=-\frac{1}{8\pi G_{0}}\left[6\tilde{\beta}^3\left(\frac{a(t)}{a(0)}\right)^{-3}\right], 
~~~\dot{P}_{BD}=\frac{1}{8\pi G_{0}}\left[2\tilde{\beta}^3\left(\frac{a(t)}{a(0)}\right)^{-3}\right]. 
\end{eqnarray}
Therefore, the speed of sound in this BD scalar field fluid turns out to be
\begin{eqnarray}
c^2_{s} = \frac{dP_{BD}}{d\rho_{BD}} = \frac{dP_{BD}/dt}{d\rho_{BD}/dt} = \frac{2\tilde{\beta}^3}
{-6\tilde{\beta}^3} = -\frac{1}{3}
\end{eqnarray}
where we followed the spirit of definition for the effective sound speed of perturbations in k-inflation
suggested by Garriga and Mukhanov \cite{mukhanov}.
Namely, it is {\it negative} and essentially the same as the equation of state above. The negative definite
sound speed squared (or the imaginary value of sound speed) may seem pathological. 
For rather an exotic type of matter like the k-essence, however, it is not so surprising to
have negative sound speed squared as it may well result from the negative pressure. Indeed, it is known
that when attempting to build a model for an accelerating universe with a barotropic perfect fluid, 
one usually runs into the problem of instabilities on short scales due to a negative sound speed
squared for the perturbations \cite{sound}. 
Therefore, the present BD theory as a k-essence model seems to suffer from the same problem
as it is a theory of barotropic fluid with negative-$w$. In this regard, we note that there are
certain ways to avoid the negative definite sound speed squared when encountered discussed 
in the literature \cite{sound}.  \\
Thus far, we have determined the behaviors 
$a(t) \sim t$, $\rho_{BD} \sim 1/a^{2}(t)$ and $P_{BD} = -\rho_{BD}/3$.  
Based on this observation, we now discuss the role played by the existence of this scalar field-dominated
era. And to this end, note first that regardless of the type of matter (i.e., whether
it is the ordinary ones like radiation and dust or the k-essence such as the BD scalar field), one can easily
realize that the strong energy condition can be represented by
\begin{eqnarray}
R_{\mu\nu}\xi^{\mu}\xi^{\nu} = 8\pi G_{0}[T_{\mu\nu} - \frac{1}{2}g_{\mu\nu}T^{\lambda}_{\lambda}]
\xi^{\mu}\xi^{\nu} = 4\pi G_{0}[3P + \rho ] \geq 0 
\end{eqnarray}
where $\xi^{\mu}$ denotes the future-directed timelike unit normal vector and we used
$\rho = T_{\mu\nu}\xi^{\mu}\xi^{\nu}$ and $T^{\lambda}_{\lambda} = (3P - \rho )$.
Next, in the standard cosmology in which ordinary type of matter (radiation or dust) is coupled to
Einstein gravity, the time-time component of the Einstein field equation gives
\begin{eqnarray}
\left(\frac{\ddot{a}}{a}\right) = - \frac{4\pi G_{0}}{3}(3P + \rho )
\end{eqnarray}
whereas in the present case of k-essence model in which the BD scalar field plays the role of the
k-essence, the time-time component of the metric field equation reads
\begin{eqnarray}
\left(\frac{\ddot{a}}{a}\right) = - \frac{\omega }{3}\left(\frac{\dot{\Phi}}{\Phi}\right)^2 -
\frac{1}{3}\left(\frac{\ddot{\Phi}}{\Phi}\right).
\end{eqnarray}
Now from eq.(9) above, we have
\begin{eqnarray}
(3P_{BD} + \rho_{BD}) = \frac{1}{16\pi G_{0}}\left[4\omega \left(\frac{\dot{\Phi}}{\Phi}\right)^2 +
6\left(\frac{\dot{a}}{a}\right)\left(\frac{\dot{\Phi}}{\Phi}\right) + 
6\left(\frac{\ddot{\Phi}}{\Phi}\right)\right].
\end{eqnarray}
Then using eq.(10) and by substituting the set of values $\omega = -3/2$, $n= - 2$ determined above,
the eqs.(18) and (19) yield
\begin{eqnarray}
\left(\frac{\ddot{a}}{a}\right) = - \frac{8\pi G_{0}}{9}(3P_{BD} + \rho_{BD} ).
\end{eqnarray}
Thus to summarize, for both the ordinary matter case and the k-essence case, the strong energy condition is always
represented by the factor $(3P + \rho )$ and it is its sign which always determines whether the universe 
decelerates or accelerates. With this preparation, we now recall the physical quantities characterizing
each era in the universe evolution which are summarized in TABLE I. 
\begin{table}
\centering
\begin{tabular}{|c|c|c|} \hline
$\rho_{rad}\sim \frac{1}{a^{4}(t)}$, $P_{rad}=\frac{1}{3}\rho_{rad}$ & 
$a(t) \sim t^{1/2}$, $(3P_{rad}+\rho_{rad})>0$ & deceleration  \\ \hline
$\rho_{m}\sim \frac{1}{a^{3}(t)}$, $P_{m}=0$ & 
$a(t) \sim t^{2/3}$, $(3P_{m}+\rho_{m})>0$ & deceleration  \\ \hline
$\rho_{BD}\sim \frac{1}{a^{2}(t)}$, $P_{BD}=-\frac{1}{3}\rho_{BD}$ & 
$a(t) \sim t $, $(3P_{BD}+\rho_{BD})=0$ & zero acceleration  \\ \hline
$\rho_{acc}=const.$, $P_{acc}= w\rho_{acc}$  $(-1\leq w< -1/3)$ & 
$a(t) \sim t^{m}$ $(m>1)$, $(3P_{acc}+\rho_{acc})<0$ & acceleration  \\ \hline
\end{tabular}
\caption{Summary of cosmic evolution in the presence of the scalar field-dominated era.} \label{t3}
\end{table}
Of course, the {\it total} energy density $\rho$ and the pressure
$P$ consist of all the contributions coming from each components, radiation, dust, BD scalar field and
some unknown entity (denoted by the subscript ``$acc$'') leading to the acceleration, namely
$\rho = \rho_{rad} + \rho_{m} + \rho_{BD} + \rho_{acc}$ and
$P = P_{rad} + P_{m} + P_{BD} + P_{acc}$
with $P_{m}=0$ as it is approximated as a dust. 
The presence of the last (i.e., the most recent) era, namely the acceleration phase has been
introduced based on the cosmological observation of the present large scale structure such as the 
anisotropy in CMBR and the luminosity of type Ia supernovae at deep redshift which all suggest that
the universe is currently undergoing cosmic acceleration and is dominated by dark energy component
with negative pressure \cite{science}. And we placed the (BD) scalar field-dominated era in between
the matter-dominated and the accelerating eras based on the way $\rho_{BD}$ scales with
the scale factor $a(t)$, namely $\rho_{BD} \sim 1/a^2(t)$. Thus with this new era being inserted, it is 
interesting to realize that now we are witnessing a grand picture of universe evolution. 
Namely to summarize,
as the universe expands, or as time goes on, the energy density $\rho $ dilutes more and more slowly, 
the pressure $P$ keeps decreasing from positive value eventually toward negative one, the scale factor
grows more and more rapidly and lastly, the strong energy condition (i.e., the sign of 
$(3P+\rho )$) moves from ``yes'' towards ``no'' leading to the transit of cosmic evolution from
{\it deceleration} towards {\it acceleration} past zero acceleration during the newly-inserted
(BD) scalar field-dominated era.  Among others, therefore, it appears that the newly found presence of
the (BD) scalar field-dominated epoch provides a picture of smooth transition from the decelerating 
matter-dominated era to the epoch of current acceleration as the acceleration is {\it zero} during this
era. Moreover, the BD scalar field itself is known to possess a generic {\it dark} nature as it is not
allowed to have direct interactions (couplings) with ordinary matter (radiation + dust) from the outset,
namely at the Largangian level since otherwise it would violate the cherished equivalence principle
as has been originally pointed out by Brans and Dicke themselves. \\ 
Indeed, one of the features of the present
study (in which the BD scalar field is taken to play the role of a k-essence field) that distinguishes
it from other k-essence models particularly aiming at the construction of the unified model for dark matter 
and dark energy can be summarized as follows. It provides no {\it direct} mechanism for the arrival
(or emergence) of the present cosmic acceleration phase. Instead, it presents a good reason or evidence
explaining why the matter-dominated era is to be followed by an accelerating phase. Indeed, if it were
not for the presence of the (BD) scalar field-dominated era (in which the
strong energy condition is on the verge of being violated, $(3P_{BD} + \rho_{BD}) = 0$ and hence
the acceleration is zero (i.e., neither deceleration nor acceleration)), the smooth, natural transition
from the {\it decelerating} matter-dominated to the {\it accelerating} era would not be achieved.
A rigorous demonstration supporting this interesting role played by the presence of the BD scalar
field-dominated era shall be given in the next section.

\begin{center}
{\rm\bf III. Effects of BD scalar field (a k-essence) on the late-time universe evolution}
\end{center}

Thus far we have ignored the possible contributions from other types of matter and concentrated on the
role played by the BD scalar field, i.e., the k-essence in order particularly to determine the behaviors
of the scale factor, energy density, pressure, and hence equation of state, etc. in the (BD) scalar
field-dominated era when the contribution from the k-essence overwhelms all others. And as a result, we
found that such (BD) scalar field-dominated era is a yet-unknown {\it zero acceleration} epoch that
should be inserted in between the decelerating matter-dominated and the accelerating eras acting as a
``crossing bridge'' between the two. Upon realizing this, then, our next mission should be a closer 
study of the effects of the k-essence (i.e., the BD scalar
field) on the evolutionary behavior of the matter-dominated and accelerating eras. Thus we now should
include all the three components, the dust-like matter, BD scalar field and the entity that drives the
current cosmic acceleration. For the sake of definiteness of our demonstration, this entity responsible
for the cosmic acceleration shall be taken as the cosmological constant $\Lambda $. Alternatively,
this study of the effects of BD scalar field on the nature of matter-dominated and accelerating eras
can be thought of as our attempt to build a ``unified model'' for dark matter-dark energy that is
currently under intensive exploration in the theoretical cosmology. In practice, 
however, the two phases, matter-dominated era and accelerating phase, cannot be analyzed in a single
framework as they have distinct equations of state. Thus in what follows, we shall consider the two
stages ; first, the matter-to-scalar field-dominated era (MATTER-TO-SCALAR) transition period and 
second, the scalar field-dominated-to-accelerating phase (SCALAR-TO-ACCELERATION) transition period. 
Particularly, the study of
the second stage in terms of the two essential ingredients, the BD scalar field $\Phi $ and the cosmological
constant $\Lambda $, can be thought of as our proposal for a k-essence model for dark energy. \\
We now reconsider the Brans-Dicke theory of gravity, this time in the presence of the
cosmological constant $\Lambda $ with mass dimension 4 as well as the dust-like matter (denoted by $m$). 
And as usual, we work in the spatially-flat FRW metric
given in eq.(4) in terms of which the Friedmann equation representing the Einstein equation and 
the Euler-Lagrange's equation of motion for the BD scalar field are given, respectively
\begin{eqnarray}
\left(\frac{\dot{a}}{a}\right)^2 &=& \frac{8\pi }{3\Phi }(\rho_{m}+\Lambda ) 
+ \frac{\omega}{6}\left(\frac{\dot{\Phi}}{\Phi}\right)^2
- \left(\frac{\dot{a}}{a}\right)\left(\frac{\dot{\Phi}}{\Phi}\right), \\
\ddot{\Phi} &+& 3\left(\frac{\dot{a}}{a}\right)\dot{\Phi} = \frac{8\pi }{(2\omega +3)}
\left[(\rho_{m}-3P_{m}) + 4\Lambda \right].
\end{eqnarray}
Once again, we stress that these field equations have to be supplemented by the consistency conditions.
Namely, in addition to these classical field equations for the metric $g_{\mu\nu}$ and the BD scalar
field $\Phi $, there is one more set of equations which are consistency conditions as they result from
the {\it geometric} Bianchi identity $\nabla_{\nu}G^{\mu\nu} = 0$. Thus from the Einstein equation we have
\begin{eqnarray}
0 = \nabla_{\nu}(R^{\mu \nu} - {1\over 2}g^{\mu \nu}R) = \nabla_{\nu}
\left[-{8\pi \over \Phi}\Lambda g^{\mu \nu} + {8\pi \over \Phi}T^{\mu \nu}_{m} + 8\pi T^{\mu \nu}_{BD}\right].
\end{eqnarray} 
Note first that, according to the original spirit of Brans and Dicke \cite{bd}, in order not to interfere
with the successes of the equivalence principle, the BD scalar field $\Phi $ is assumed not to enter
into the equations of motion of ordinary (dust) matter so that $T^{\mu\nu}_{m}$ obeys
the usual conservation law $\nabla_{\nu}T^{\mu\nu}_{m}=0$ which, in terms of the FRW metric, takes the
familiar form
\begin{eqnarray}
\dot{\rho}_{m} = -3\left(\frac{\dot{a}}{a}\right)(\rho_{m}+P_{m})
\end{eqnarray}
where again $P_{m}=0$ in dust approximation. Therefore, we are left with
\begin{eqnarray}
0 = 8\pi \Lambda \frac{(\partial_{\nu}\Phi)}{\Phi^2}g^{\mu\nu} - 8\pi \frac{(\partial_{\nu}\Phi)}{\Phi^2}T^{\mu\nu}_{m}
+ 8\pi G_{0}\nabla_{\nu}[P_{BD}g^{\mu\nu} + (\rho_{BD}+P_{BD})U^{\mu}U^{\nu}]
\end{eqnarray}
which, again in terms of the FRW metric, becomes
\begin{eqnarray}
\dot{\rho}_{BD} + 3\left(\frac{\dot{a}}{a}\right)(\rho_{BD}+P_{BD}) = \frac{1}{G_{0}}
(\rho_{m}+\Lambda )\left(\frac{\dot{\Phi}}{\Phi^2}\right).
\end{eqnarray}
Thus in order to determine the natute of late-time universe evolution, one in principle has to attemt to 
solve the coupled, non-linear field equations given in eqs.(21) and (22) subject to the consistency 
conditions in eqs.(24) and (26).  
\\
{\bf 1. MATTER-TO-SCALAR transition period}
\\
At this stage, the energy density of universe is assumed to be dominated by those of dust-like matter,
$\rho_{m}$ and BD scalar field, $\rho_{BD}$. First from the energy-momentum conservation equation (24), 
the matter energy density $\rho_{m}$ is given by
\begin{eqnarray}
\rho_{m} = \frac{1}{G^2_{0}}\left[\frac{a(t)}{a(0)}\right]^{-3}.
\end{eqnarray}
Then the coupled field equations for this transition period becomes
\begin{eqnarray}
\left(\frac{\dot{a}}{a}\right)^2 &=& \frac{8\pi }{3G^2_{0} }\frac{\kappa}{\Phi a^3} 
+ \frac{\omega}{6}\left(\frac{\dot{\Phi}}{\Phi}\right)^2
- \left(\frac{\dot{a}}{a}\right)\left(\frac{\dot{\Phi}}{\Phi}\right), \\
\ddot{\Phi} &+& 3\left(\frac{\dot{a}}{a}\right)\dot{\Phi} = \frac{8\pi \kappa}{(2\omega +3)G^2_{0}}
\frac{1}{a^3}
\end{eqnarray}
where $\kappa \equiv a^3(0)$
and of course these field equations have to be supplemented by the remaining consistency equation
\begin{eqnarray}
\dot{\rho}_{BD} + 3\left(\frac{\dot{a}}{a}\right)(\rho_{BD}+P_{BD}) = \frac{\kappa}{G^3_{0}}
\frac{1}{a^3}\left(\frac{\dot{\Phi}}{\Phi^2}\right).
\end{eqnarray}
We now start with the solution ansatz
\begin{eqnarray}
\Phi(t) = \frac{1}{G_{0}}\left[1 + \chi t\right]^{\gamma}, 
~~~a(t) = a(0)\left[1 + \chi t\right]^{\alpha}.
\end{eqnarray}
Substituting these solution ansatz into the coupled field equations (28) and (29) above yields,
\begin{eqnarray}
\alpha = \frac{2(\omega +1)}{(3\omega +4)}, ~~~\gamma = \frac{2}{(3\omega +4)},
~~~\chi^2 = \frac{4\pi}{G_{0}}\frac{(3\omega +4)}{(2\omega +3)}.
\end{eqnarray}
Thus basically the solutions behave as
\begin{eqnarray}
\Phi(t) \sim t^{2/(3\omega +4)}, ~~~a(t) \sim t^{2(\omega +1)/(3\omega +4)}
\end{eqnarray}
and they have actually been known \cite{weinberg} for some time in the Brans-Dicke cosmology.
Note that in the standard Einstein gravity limit $\omega \to \infty$ in which the dynamics of the
BD scalar field is washed out, i.e., $\Phi \to 1/G_{0}$, one recovers the standard behavior
for the matter-dominated era, $a(t) \sim t^{2/3}$. 
With these exact solutions at hand, next we consider the behavior of the BD scalar field
(being treated as a perfect fluid) in this transition period. Recall, first that
its energy density and pressure are given by eq.(9) earlier, which have been obtained using eqs.(2)
and (8). Thus by plugging the exact solutions for the case at hand, eqs.(31) and (32) into eq.(9),
we end up with
\begin{eqnarray}
\rho_{BD} &=& -\frac{1}{G^{2}_{0}}\left[\frac{(5\omega +6)}{(2\omega +3)(3\omega +4)} 
\left(\frac{a(t)}{a(0)}\right)^{-2/\alpha}\right],  \\
P_{BD} &=& \frac{1}{G^{2}_{0}}\left[\frac{2(\omega +1)}{(2\omega +3)(3\omega +4)} 
\left(\frac{a(t)}{a(0)}\right)^{-2/\alpha}\right]  \nonumber
\end{eqnarray}
where $\alpha$ is as given in eq.(32).
Here a causion needs to be exercised. Namely, these expressions for $\rho_{BD}$ and $P_{BD}$ do not
necessarily mean that $\rho_{BD}<0$ and $P_{BD}>0$. Indeed, it is the other way around if we demand
the positive-definite energy density for the BD scalar field, i.e., $\rho_{BD}>0$. Then it immediately
follows that $P_{BD}<0$. To see this is indeed the case, note that demanding $\rho_{BD}>0$ leads to
$\omega <-3/2$ or $-4/3<\omega <-6/5$ which, in turn, indicates $(2\omega +3)(3\omega +4)>0$ and 
$(\omega +1)<0$ and also $(5\omega +6)<0$. Therefore, clearly $P_{BD}<0$. Besides, this condition
$\rho_{BD} > 0$ (particularly $\omega <-3/2$) also guarantees the condition of cosmic expansion
$\alpha = 2(\omega +1)/(3\omega +4) > 0$. Next note that as a result,
the equation of state of the BD scalar field perfect fluid in this transition period is given by
\begin{eqnarray}
w_{BD} = \frac{P_{BD}}{\rho_{BD}} = -\frac{2(\omega +1)}{(5\omega +6)} < 0.
\end{eqnarray}
Namely, it is {\it negative} implying that already in this matter-to-scalar field-dominated era transition 
period, the BD scalar field behaves as a negative pressure component. In the earlier section in which
we maintained the BD scalar field alone and ignored the contributions from other types of matter, it
has been realized that the equation of state of the BD scalar field is $w_{BD} = P_{BD}/\rho_{BD}
= -1/3$ implying the emergence of zero acceleration epoch driven by the negative pressure. This
negative pressure nature of the BD scalar field as a k-essence, therefore, appears to begin already in
this transition period. Lastly, note that these solutions in eqs.(31) and (32) indeed satisfy the
remaining consistency condition in eq.(30). This consistency condition, however, does not determine the
BD $\omega $-parameter to a particular value.
\\
{\bf 2. SCALAR-TO-ACCELERATION transition period}
\\
At this stage, the universe energy density is assumed to be dominated by that of BD scalar field, 
$\rho_{BD}$ and the cosmological constant $\Lambda $.
And as mentioned earlier, the study of this second stage involving essentially the two components, 
the BD scalar field $\Phi $ and the cosmological constant $\Lambda $, can be thought of as our proposal 
for a k-essence model for dark energy. Then the difference between our dark energy model and those of other
quintessence/k-essence theories lies in the fact that here in our model, we are interested in the way
the presence of the BD scalar field, i.e., a k-essence, modifies the evolutionary behavior of the
vacuum energy ($\Lambda$)-dominated epoch (which has an exponentially expanding nature in the standard general
relativity context) while there, the quintessence/k-essence fields themselves (without the $\Lambda$-term)
are expected to generate an accelerating expansion of some sort. \\ 
The coupled Friedmann equation and the BD scalar field equation for the case at hand amount to setting 
$\rho_{m} = 0$ in eqs.(21) and (22), namely
\begin{eqnarray}
\left(\frac{\dot{a}}{a}\right)^2 &=& \frac{8\pi }{3\Phi}\Lambda 
+ \frac{\omega}{6}\left(\frac{\dot{\Phi}}{\Phi}\right)^2
- \left(\frac{\dot{a}}{a}\right)\left(\frac{\dot{\Phi}}{\Phi}\right), \\
\ddot{\Phi} &+& 3\left(\frac{\dot{a}}{a}\right)\dot{\Phi} = \frac{8\pi}{(2\omega +3)}
4\Lambda.
\end{eqnarray}
Again, these field equations should be supplemented by the remaining consistency equation
\begin{eqnarray}
\dot{\rho}_{BD} + 3\left(\frac{\dot{a}}{a}\right)(\rho_{BD}+P_{BD}) = \frac{\Lambda}{G_{0}}
\left(\frac{\dot{\Phi}}{\Phi^2}\right).
\end{eqnarray}
Again, we begin with the solution ansatz
\begin{eqnarray}
\Phi(t) = \frac{1}{G_{0}}\left[1 + \chi t\right]^{\gamma}, 
~~~a(t) = a(0)\left[1 + \chi t\right]^{\alpha}.
\end{eqnarray}
Substituting these solution ansatz into the coupled field equations (36) and (37) yields in this time,
\begin{eqnarray}
\alpha &=& \frac{(2\omega +1)}{2},  ~~~\gamma = 2, \\
\chi &=& \tilde{\chi}/h, ~~~\tilde{\chi}^2 = \frac{8\pi G_{0}}{3}\Lambda,
~~~h^2 = \frac{(2\omega +3)(6\omega +5)}{12}.  \nonumber
\end{eqnarray}
Thus the solutions behave as
\begin{eqnarray}
\Phi(t) = \frac{1}{G_{0}}\left[1 + \chi t\right]^{2}, 
~~~a(t) = a(0)\left[1 + \chi t\right]^{(2\omega +1)/2}.  \nonumber
\end{eqnarray}
Having these exact solutions with us, again we consider the behavior of the BD scalar field
(i.e., a k-essence) in this transition period.  
By plugging the exact solutions for the case at hand, eqs.(39) and (40) into eq.(9),
we are left with 
\begin{eqnarray}
\rho_{BD} &=& -\Lambda \left[\frac{4(4\omega +3)}{(2\omega +3)(6\omega +5)} 
\left(\frac{a(t)}{a(0)}\right)^{-2/\alpha}\right],  \\
P_{BD} &=& \Lambda \left[\frac{8(3\omega +2)}{(2\omega +3)(6\omega +5)} 
\left(\frac{a(t)}{a(0)}\right)^{-2/\alpha}\right] \nonumber
\end{eqnarray}
with $\alpha$ being given in eq.(40) and as a result,
the equation of state of the BD scalar field perfect fluid in this transition period is given by
\begin{eqnarray}
w_{BD} = \frac{P_{BD}}{\rho_{BD}} = -\frac{2(3\omega +2)}{(4\omega +3)} < 0.
\end{eqnarray}
In this transition period, however, the cosmic acceleration is supposed to set in. Thus one should
demand $\alpha = (2\omega +1)/2 > 1$ or $\omega > 1/2$ and this, in turn, leads to $\rho_{BD}<0$ and
$P_{BD}>0$ as $\Lambda > 0$ from eq.(40). Namely at this stage, the BD scalar field turns around and
begins to behave as a {\it positive} pressure component making the total pressure 
$P_{tot} = P_{\Lambda} + P_{BD} = -\Lambda + P_{BD}$ {\it less} negative and hence {\it slowing down}
the accelerated expansion rate from the exponential law to the power law as we shall discuss in more
detail below. Note, however, that since $a(t) \sim t^{\alpha}$ and hence 
$\rho_{BD}, P_{BD} \sim a^{-2/\alpha}(t) \sim 1/t^{2}$, the {\it positive} contribution of $P_{BD}$
to $P_{tot} = -\Lambda + P_{BD}$ would eventually become negligible as time goes on and the same
argument holds for the total energy density $\rho_{tot} = \rho_{\Lambda} + \rho_{BD} = \Lambda + \rho_{BD}$
namely, the {\it negative} contribution of $\rho_{BD}$ to $\rho_{tot}$ would become more and more
negligible. This last observation, then, appears to indicate that the accelerated expansion would
eventually ``speed up''.
Lastly, note that these solutions in eqs.(39) and (40) indeed satisfy the remaining consistency condition
in eq.(38). Again, this consistency condition
does not determine the BD $\omega$-parameter to a particular value.  \\
These solutions in eqs.(39) and (40) have indeed been found some time ago in the so-called 
{\it extended inflation} model \cite{la} which attempts to resolve the {\it graceful exit} problem and
hence to save the original spirit of ``old inflation'' scenario. 
\\
{\bf 3. Remarks}
\\
(I) We go back to the case of MATTER-TO-SCALAR transition period studied above and examine
the nature of solutions there. First, the condition of ``cosmic expansion'' amounts to demanding
$\alpha = 2(\omega +1)/(3\omega +4) >0$, which yields $\omega <-4/3$ or $\omega >-1$. Secondly, 
there we demanded $\rho_{BD}>0$ which led to $\omega <-3/2$ or $-4/3<\omega <-6/5$.
Thus these two conditions, when combined, give $\omega <-3/2$. This, in turn, allows us to conclude that
\begin{eqnarray}
\left(\alpha - \frac{2}{3}\right) = \frac{-2}{3(3\omega +4)} > 0. 
\end{eqnarray}
This inequality indicates that the expansion rate of the matter-dominated era in the presence of the
BD scalar field, which is a k-essence, (or equivalently in the context of BD gravity theory) namely,
$a(t)\sim t^{\alpha}$, turns out to be greater than that, i.e., $a(t)\sim t^{2/3}$ in the absence of 
the k-essence field (or in the context of general relativity). 
Note also that from eqs.(27), (31) and (34), 
$\rho_{m} \sim a^{-3}(t) \sim 1/t^{3\alpha} < \rho_{BD} \sim a^{-2/\alpha}(t) \sim 1/t^2$ as 
$\alpha > 2/3$. Recall from section II that $a(t) \sim t$ and thus $\rho_{BD} \sim a^{-2}(t) \sim 1/t^2$
in the BD scalar field-dominated era. This observation indicates that as the evolution proceeds,
the BD scalar field component takes over the energy density of the matter-dominated era and as a result, 
the BD scalar field-dominated era does indeed arrive as we assumed in the section II. And this supports
our suggestion for the grand picture of universe evolution summarized in TABLE I. \\
(II) Next, in the case of SCALAR-TO-ACCELERATION transition period, we learned that the accelerated 
expansion rate in the vacuum energy ($\Lambda$)-dominated era slows down to a power law by introducing
the BD scalar field, which is again a k-essence (or equivalently in the context of BD theory). \\
Now these two observations (I) and (II) appear to imply that the effects of the presence of the BD scalar
field (playing the role of an unique k-essence) on the late-time evolution of the universe can be
summarized as follows ; the presence of the k-essence appears to interpolate {\it smoothly} between the 
decelerating matter-dominated era and the accelerating phase by speeding up the expansion rate of the 
matter-dominated era somewhat while slowing down the expansion rate of the late-time cosmic
acceleration phase to some extent ! This effect coming from the presence of the BD scalar field has 
been summarized in TABLE II. 
\begin{table}
\centering
\begin{tabular}{|c|c|c|} \hline
       & {\rm General Relativity (without BD scalar)} & {\rm Brans-Dicke Theory (with BD scalar)} \\ \hline
{\rm Matter-dominated era} & $a(t) \sim t^{2/3}$ &
$a(t) \sim t^{\alpha}$, $(\alpha > 2/3)$ \\ \hline
{\rm Accelerating phase} & $a(t) = a(0)e^{\tilde{\chi}t}$ &
$a(t) = a(0)[1+\chi t]^{(2\omega +1)/2}$  \\ \hline
\end{tabular}
\caption{Effect of the presence of the BD scalar field on the universe's late-time evolution.} \label{t2}
\end{table}
And this role of ``crossing bridge'' between the two late-time epochs is indeed consistent with the result of 
our earlier study of BD scalar field as a unique k-essence in which we idealized the situation by ignoring 
all other types of matter, namely, the emergence of a zero-acceleration epoch in between the two late-time
eras. \\
(III) Lastly, it seems worthy of note that the BD parameter $\omega $ ``runs'' with the scale in order to
serve as a successful model for dark matter - dark energy. Namely, it appears to be a {\it running}
coupling parameter with {\it growing} behavior
\begin{eqnarray}
\left\{%
\begin{array}{ll}
 \omega <-3/2 ~({\rm say}) & \hbox{{\rm in MATTER-TO-SCALAR transition period}, } \\
 \omega = -3/2 & \hbox{{\rm in (BD) scalar field-dominated era},} \\
 \omega > 1/2 & \hbox{{\rm in SCALAR-TO-ACCELERATION transition period}} \\
\end{array}%
\right.
\nonumber
\end{eqnarray}
as the scale factor $a(t)$ increases (with time). As we mentioned earlier, the BD parameter $\omega $ in
our model is a parameter of a k-essence (i.e., the BD scalar field) theory. Thus from a scalar field
theory perspective, this ``running'' behavior of the BD $\omega $-parameter can be viewed as being
natural in the sense of the renormalization group approach. What seems to be noteworthy is the fact that 
its value appears to grow with growing scale factor.

\begin{center}
{\rm\bf IV. Concluding remarks}
\end{center}

We begin with the comparison of our BD scalar field-dominated universe studied in section II with
the curvature-dominated universe. As is well-known, the curvature-dominated universe is (effectively)
an ``empty space'' solution in the spatially-open ($k=-1$) FRW model of Einstein gravity. As such,
the universe expansion occurs essentially due to the {\it negative} curvature of the spatial 
section ($k=-1$). That is, all types of matter are absent (or assumed to be negligible) and the
curvature term in the Friedmann equation behaves as if it is a some kind of energy density term,
$(\dot{a}/a)^2 = -k/a^2 = 1/a^2$ which gives $a(t)\sim t$. And the acceleration is zero simply
because $P=0=\rho $ in eq.(17), i.e., $(\ddot{a}/a) = -(4\pi G_{0}/3)(3P+\rho ) = 0$. \\
By contrast, the BD scalar field-dominated universe can be thought of as the k-essence-dominated
solution in the spatially-flat ($k=0$) FRW model of a non-minimally coupled Einstein-scalar theory
(or the pure BD theory of gravity). And the universe expansion occurs due to the k-essence
(i.e., the BD scalar field) energy density $\rho_{BD}\sim 1/a^2$ in the Friedmann equation,
$(\dot{a}/a)^2 = (8\pi G_{0}/3)\rho_{BD} = \tilde{\beta}^2a^2(0)/a^2$ (where we used eqs.(5),(9) and
(13)). Particularly the zero acceleration can be attributed to the {\it negative} pressure
$P_{BD}=-\rho_{BD}/3$ in eq.(20), i.e., $(\ddot{a}/a) = -(8\pi G_{0}/9)(3P_{BD}+\rho_{BD}) = 0$. 
To summarize, although the two universe models appear to exhibit the same expansion behavior
$a(t)\sim t$, $\ddot{a} = 0$, the origin/nature of zero acceleration of the former comes from the
{\it negative} spatial curvature (a geometrical nature) while that of the latter comes from the
{\it negative} pressure of the BD scalar field or k-essence (a matter).  Of course the essential
difference between the two arises from the fact that the spatial section of the universe is open for
the former model whereas it is flat for the latter model. \\  
Next, it would be interesting to contrast in some more detail the way the BD scalar field plays a role of 
a k-essence with that other 
k-essence candidates do concerning the unified dark matter - dark energy model.
First, in the generalized Chaplygin gas model \cite{chaplygin}, the equation of state and the energy 
density are given by
\begin{eqnarray}
P = - \frac{A}{\rho^{\alpha}},
~~~\rho = \left[A + \frac{B}{a^{3(1+\alpha)}(t)}\right]^{1/(1+\alpha)}. 
\end{eqnarray}
Note that the original
Chaplygin gas model amounts to the case $\alpha = 1$. And in this generalized Chaplygin gas model,
the sound speed is small at early times and becomes larger at late times. Next, in the so-called, 
``purely kinetic'' k-essence model recently studied by Scherrer \cite{scherrer}, the energy density 
of the associated k-essence field reads
\begin{eqnarray}
\rho = \rho_{0} + \rho_{1}\left(\frac{a(t)}{a_{0}}\right)^{-3}
\end{eqnarray}
where $\rho_{0}$ and $\rho_{1}$ are some constants and $a_{0}$ denotes the value of scale factor today. 
In this model of Scherrer, the speed of sound is 
small at all times. This last point has been emphasized as an advantage of this model over the previous
(generalized) Chaplygin gas model in which the sound speed becomes larger at late times like present
epoch which is certainly an unwanted feature. Both in these two models, the energy density $\rho $ 
behaves as that of matter at early times (when the scale factor $a(t)$ is small) while at late times
it acts like a cosmological constant. In this respect, these two classes of theories have been 
considered to be able to serve as unified models for dark matter and dark energy.   
On the other hand, although the present study
of BD theory as a k-essence model has little to say about the nature of emergence of the cosmic
acceleration phase, it instead provides a crossing bridge between the matter-dominated and the
accelerating eras as it is an intermediate regime of zero acceleration. 
We find this as certainly a remarkable feature that distinguishes our model from 
other k-essence models in the literaure that we briefly described above. 
Lastly, one may wonder if the presence of an additional epoch, i.e., the (BD) scalar field-dominated
era, could be inconsistent with the current estimate (and observation by WMAP \cite{wmap}) of the age of
the universe based on the assumption that the universe is presently matter-dominated and should have been 
so for most of its history. Under this assumption, using $H_{0} = \dot{a}/a = \alpha /t$, 
($\alpha > 2/3$) for the 
matter-domination and its observed value $H_{0} \simeq 70 km ~s^{-1} ~Mpc^{-1}$, the age of the universe is
estimated to be $\tau \simeq 14 ~(Gyr)$. However, some parts of the late time that have been 
regarded as being matter-dominated so far should now be replaced by the scalar field-dominated and 
accelerating phases. Therefore, durations of all the three eras
(allowing, of course, for their possible overlaps) would instead sum up to the age of the universe, 
which is $14 ~(Gyr)$. \\
This work was financially supported by the BK21 Project of the Korean Government.

\end{document}